\documentclass[
reprint,           
superscriptaddress,
amsmath,           
amssymb,           
aps,               
prl,               
notitlepage,       
floatfix,          
nofootinbib,
onecolumn,
]{revtex4-1}

\usepackage{tensor}     
\usepackage{graphicx}   
\usepackage[
colorlinks=true,        
citecolor=blue,         
linkcolor=blue,         
urlcolor=blue           
]{hyperref}             
\usepackage{bm}         
\usepackage{xcolor}     
\usepackage{tabulary}
\usepackage{color}      
\usepackage[utf8]{inputenc} 
\usepackage[section]{placeins} 

\newcommand{\nc}{\newcommand*} 

\nc{\al}{\alpha}
\nc{\s}{\sigma}
\nc{\dt}{\delta}
\nc{\Dt}{\Delta}
\nc{\Ld}{\Lambda}
\nc{\p}{\partial}
\nc{\om}{\omega}
\nc{\Om}{\Omega}
\nc{\rd}{\mathrm{d}}
\nc{\Od}[1]{\mathcal{O}(#1)} 
\nc{\kp}{\kappa}
\nc{\one}{\uppercase\expandafter{\romannumeral1}}
\nc{\two}{\uppercase\expandafter{\romannumeral2}}
\nc{\three}{\uppercase\expandafter{\romannumeral3}}
\def\({\left(}
\def\){\right)}
\def\[{\left[}
\def\]{\right]}
\def\e{\begin{equation}}
\def\q{\end{equation}}
\def\m{\begin{eqnarray}}
\def\n{\end{eqnarray}}
\nc{\Eq}[1]{Eq.~\eqref{#1}}     
\nc{\Fig}[1]{Fig.~\ref{#1}}     
\nc{\Table}[1]{Table~\ref{#1}}  
\nc{\Sec}[1]{Sec.~\ref{#1}}     
\nc{\Msun}{M_\odot}             
\nc{\fpbh}{f_{\mathrm{pbh}}}    
\nc{\fpbhn}{f_{\mathrm{pbh0}}}    
\nc{\mR}{\mathcal{R}} 
\nc{\seq}{\sigma_{\mathrm{eq}}}
\nc{\ogw}{\Omega_{\mathrm{GW}}}
\nc{\gpcyr}{\mathrm{Gpc}^{-3}\,\mathrm{yr}^{-1}}
\nc{\lvc}{LIGO/Virgo} 
\nc{\SNR}{\mathrm{SNR}} 
\nc{\mmin}{{m_{\mathrm{min}}}}
\nc{\mmax}{{m_{\mathrm{max}}}}
\nc{\Mmin}{{M_{\mathrm{min}}}}
\nc{\fmin}{{f_{\mathrm{min}}}}
\nc{\VT}{\mathrm{VT}}
\nc{\rhoGW}{\rho_{\mathrm{GW}}}
\nc{\vth}{\vec{\theta}}
\nc{\vd}{\vec{d}}
\nc{\vla}{\vec{\lambda}}
\nc{\Nobs}{N_{\mathrm{obs}}}
\nc{\av}[1]{\langle #1 \rangle} 
\nc{\km}{\mathrm{km}}
\nc{\Mpc}{\mathrm{Mpc}}
\nc{\Tobs}{T_{\mathrm{obs}}}
\nc{\Ntemp}{N_{\mathrm{temp}}}
\nc{\addref}{[\textcolor{red}{add ref}] } 
\nc{\eg}{\textit{e.g.~}}
\nc{\app}{\approx}
\nc{\hf}{\frac{1}{2}}
\nc{\discuss}{\textcolor{red}{Add discussion here!}}
\nc{\red}[1]{\textcolor{red}{#1}}
\nc{\mH}{\mathcal{H}}
\nc{\cs}{c_s^2}
\nc{\Sij}[1]{S_{ij}^{(#1)}}
\nc{\vi}[1]{v_i^{(#1)}}
\nc{\no}{\nonumber}
\def\<{\left\langle}
\def\>{\right\rangle}

\def\half{{1\over 2}}
\nc{\bk}{\bm{k}}
\nc{\bq}{\bm{q}}
\nc{\bp}{\bm{p}}
\nc{\bl}{\bm{l}}
\nc{\bx}{\bm{x}}
\nc{\be}{\mathbf{e}}
\nc{\mS}{\mathcal{S}}
\nc{\te}{\tilde{\eta}}
\nc{\tp}{\tilde{p}}
\nc{\tk}{\tilde{k}}
\nc{\tx}{\tilde{x}}
\nc{\tF}{\tilde{F}}
\nc{\tA}{\tilde{A}}
\nc{\mkpq}{|\bk-\bp-\bq|}
\nc{\mpq}{|\bp-\bq|}
\nc{\mkp}{|\bk-\bp|}
\nc{\mSi}[1]{\mS^{(#1)}({\bk, \eta})}
\nc{\vk}{\vec{k}}
\nc{\kstar}{k_*}
\nc{\xstar}{x_*}
\nc{\mpbh}{m_{\rm{pbh}}}
\nc{\Ci}{\mathrm{Ci}}
\nc{\Si}{\mathrm{Si}}

\renewcommand{\vec}[1]{\boldsymbol{#1}} 

\begin{document}
	
\title{Scalar Induced Gravitational Waves in Different Gauges}
	
\author{Chen Yuan}
\email{yuanchen@itp.ac.cn}
\affiliation{CAS Key Laboratory of Theoretical Physics, 
Institute of Theoretical Physics, Chinese Academy of Sciences,
Beijing 100190, China}
\affiliation{School of Physical Sciences, 
University of Chinese Academy of Sciences, 
No. 19A Yuquan Road, Beijing 100049, China}
	
\author{Zu-Cheng Chen}
\email{chenzucheng@itp.ac.cn} 
\affiliation{CAS Key Laboratory of Theoretical Physics, 
Institute of Theoretical Physics, Chinese Academy of Sciences,
Beijing 100190, China}
\affiliation{School of Physical Sciences, 
University of Chinese Academy of Sciences, 
No. 19A Yuquan Road, Beijing 100049, China}

\author{Qing-Guo Huang}
\email{huangqg@itp.ac.cn}
\affiliation{CAS Key Laboratory of Theoretical Physics, 
Institute of Theoretical Physics, Chinese Academy of Sciences,
Beijing 100190, China}
\affiliation{School of Physical Sciences, 
University of Chinese Academy of Sciences, 
No. 19A Yuquan Road, Beijing 100049, China}
\affiliation{Center for Gravitation and Cosmology, 
College of Physical Science and Technology, 
Yangzhou University, Yangzhou 225009, China}
\affiliation{Synergetic Innovation Center for Quantum Effects and Applications, 
Hunan Normal University, Changsha 410081, China}
	
\date{\today}

\begin{abstract}
    
In this letter we calculate the scalar induced gravitational waves (SIGWs) accompanying the formation of primordial black hole during the radiation dominated era in three different gauges, i.e. synchronous gauge, Newton gauge and uniform curvature gauge, and we find that the energy density spectra of SIGWs, $\ogw(k)$, are identical in these three different gauges.

\end{abstract}
	
\pacs{???}
	
\maketitle
	

{\it Introduction. } Primordial black holes (PBHs) have attracted a lot of attention recently, because they not only provide a possible explanation \cite{Sasaki:2016jop,Chen:2018czv} to the mergers of binary black holes discovered by LIGO \cite{Abbott:2016blz}, but also can compose of all of dark matter (DM) in a substantial window in the mass range of $[10^{-16},10^{-14}] \cup [10^{-13},10^{-12}] M_\odot$ even though there are many constraints on the abundance of PBHs in DM from various observations \cite{Tisserand:2006zx,Carr:2009jm,Barnacka:2012bm,Griest:2013esa,Graham:2015apa,Brandt:2016aco,Chen:2016pud,Wang:2016ana,Gaggero:2016dpq,Ali-Haimoud:2016mbv,Blum:2016cjs,Horowitz:2016lib,Niikura:2017zjd,Zumalacarregui:2017qqd,Abbott:2018oah,Magee:2018opb,Chen:2018rzo,Niikura:2019kqi,Chen:2019irf,Authors:2019qbw,Wang:2019kzb}.

PBHs form from the gravitational collapse of over-densed regions seeded by large curvature perturbations \cite{Hawking:1971ei,Carr:1974nx} on small scales after the corresponding wavelength re-enters the horizon once the curvature perturbations exceed a critical value. Since  the curvature perturbations couple to the tensor perturbations at second-order, such large curvature perturbations also produce the so-called scalar induced gravitational waves (SIGWs) in the radiation dominated era \cite{tomita1967non,Matarrese:1992rp,Matarrese:1993zf,Matarrese:1997ay,Noh:2004bc,Carbone:2004iv,Nakamura:2004rm}. SIGWs are supposed to be a new probe to the PBHs \cite{Yuan:2019udt,Yuan:2019wwo,Chen:2019xse}. See some other relevant papers in \cite{Ananda:2006af,Baumann:2007zm,Saito:2008jc,Assadullahi:2009jc,Bugaev:2009zh,Saito:2009jt,Bugaev:2010bb,Nakama:2016enz,Nakama:2016gzw,Garcia-Bellido:2017aan,Sasaki:2018dmp,Espinosa:2018eve,Kohri:2018awv,Cai:2018dig,Bartolo:2018evs,Bartolo:2018rku,Unal:2018yaa,Inomata:2018epa,Clesse:2018ogk,Cai:2019amo,Inomata:2019zqy,Inomata:2019ivs,Cai:2019elf,Cai:2019cdl,Tomikawa:2019tvi,DeLuca:2019ufz}.

Unfortunately, even though the gravitational waves (GWs) are gauge invariant at first order, they fail to remain so at the second order (see e.g., \cite{Noh:2003yg}).
It is therefore quite natural to suspect that the energy density spectrum of SIGWs, $\ogw(k)$, is also gauge dependent.
The SIGWs in different gauges are calculated in \cite{Hwang:2017oxa,Gong:2019mui,Tomikawa:2019tvi,DeLuca:2019ufz}. In particular, the authors in \cite{DeLuca:2019ufz} pointed out that the response of detector like LISA to
the SIGW signals should be calculated in synchronous gauge because the projected sensitivity curves for the LISA are given in that gauge.
Interestingly, they also noticed that $\ogw(k)$ in the synchronous gauge is at least one order of magnitude smaller than the one in the Newton gauge \cite{DeLuca:2019ufz}.
In this letter, we will carefully revisit the gauge problem in SIGWs by explicitly calculating $\ogw(k)$ in three different gauges: synchronous gauge, Newton gauge and uniform curvature gauge, and we find that the spectra of $\ogw(k)$ are identical in these three gauges explicitly.

{\it Perturbations of metric. } The most generic perturbed metric that contains scalar perturbations and GWs can be generally written by 
\m
g_{00}&&=-1-2\phi,\no\\
g_{0i}&&=a\partial_iB,\no\\
g_{ij}&&=a^2\delta_{ij}+a^2 \(\half h_{ij}-2\delta_{ij}\psi+2\p_i\p_jC\),
\n
where $\phi$, $B$, $\psi$ and $C$ are four scalar modes of the metric perturbation, and $h_{ij}$ is the second-order tensor mode which should satisfy $h_i^i=0$. On the other hand, the perturbed stress tensor described by perfect fluid up to first-order reads
\m
T_{00}&&=\rho+2\rho\phi+\delta\rho,\no\\
T_{0i}&&=-\rho\p_iB-P\p_i v-\rho\p_i v,\no\\
T_{ij}&&=(P+\delta P)\delta_{ij}-2P\psi\delta_{ij}+2P\p_i\p_jC,
\n
where $v$ is the velocity potential for irrotational scalar perturbations.
If one considers a first-order change in the coordinate such that
\e
\tilde{\eta}= \eta+T,~\tilde{x}^i= x^i+\p^iL,
\q
the scalar modes will transform as \cite{Malik:1998ai}
\m
\tilde{\phi}&&=\phi-\mH T-T',\\
\tilde{\psi}&&=\psi+\mH T,\\
\tilde{B}&&=B+T-L',\\
\tilde{C}&&=C-L,\\
\tilde{v}&&=v+L',
\n
where a prime denotes a derivative with respect to conformal time $\eta$.
Fixing $T$ determines the time-slicing while choosing $L$ decides the spatial coordinates used to calculate the physics quantities.
Based on these transformations, two gauge-invariant quantities can be constructed as
\m\label{Phi}
\Phi&&\equiv\phi-\mH\sigma-\sigma',\quad\Psi\equiv\psi+\mH\sigma,
\n
where $\sigma\equiv C'-B$ is the shear potential.
One can show from the first-order Einstein equation that $\Phi=\Psi$ in the absence of anisotropies. Below we will investigate the energy density spectrum of SIGWs under different gauge conditions. We perform all the calculations in a radiation dominated (RD) universe.
The evolution of the GWs is govern by
\e\label{eqh}
h_{i j}^{\prime \prime}+2 \mathcal{H} h_{i j}^{\prime}-\nabla^{2} h_{i j}=-4 \mathcal{T}_{i j}^{\ell m} S_{\ell m},
\q
where $\mH \equiv a'/a$ is the conformal Hubble parameter, and 
$\mathcal{T}_{i j}^{\ell m}$ is the projection operator \cite{Ando:2017veq} 
onto the transverse and traceless tensor.
An expression for the source term, $S_{\ell m}$, without fixing gauge can be found in \cite{Gong:2019mui}.
The solution of GWs in Fourier space can be obtained by Green's function, namely
\e
h(\eta,\vec{k})={1\over a(\eta)}\int_0^\eta g_k(\eta;\eta')a(\eta')S(\eta',\vec{k})\mathrm{d}\eta',
\q
where $S(\vec{k},\eta)\equiv -4e^{ij}(\vec{k})\mathcal{S}(k,\eta)$ and $\mathcal{S}(k,\eta)$ is the source term in Fourier space. The Green's function is given by $g_k(\eta;\eta')={1\over k}\sin(k\eta-k\eta')$ in a RD universe.

{\it SIGWs in synchronous gauge. }
The synchronous gauge, which is dubbed TT gauge in \cite{DeLuca:2019ufz}, requires that $\phi=B=0$. The corresponding source term is
\m
S_{ij}^{S}&&=-3\psi'\p_i\p_j\sigma+\psi\p_i\p_j\psi-{1\over \mH^2}\(\p_i\psi'\)\(\p_j\psi'\)+\(\p^2\sigma\)\(\p_i\p_j\sigma\)-\(\p^k\p_i\sigma\)\(\p_k\p_j\sigma\).
\n
The equation of motion for $\phi$ and $\sigma$ can be obtained from the first-order spatial Einstein equations, namely
\m
2\mH\sigma+\sigma'+\psi&&=0,\\
6\psi''+2\mH\(9\psi'-4\p^2\sigma\)-3\p^2\sigma'-5\p^2\psi&&=0.
\n
The solutions to above equations are
\m\label{psisigma}
\psi(\vec{k})&&\equiv\Phi_k
T_\psi(k\eta)=9\Phi_k{1-\cos(k\eta/\sqrt{3})\over (k\eta)^2},\\
k\sigma(\vec{k})&&\equiv
\Phi_kT_\sigma(k\eta)=9\Phi_k{\sqrt{3}\sin(k\eta/\sqrt{3})-k\eta\over (k\eta)^2},
\n
where we have normalized the coefficient such that the initial value of $\Phi$ is $\Phi_k$.

In Fourier space, the source term becomes
\m
S^{S}(\vec{k},\eta)
&&=-4\int\frac{\mathrm{d}^3p}{(2\pi)^{3/2}}\Big(e^{ij}p_ip_j\Big)\Big\{3\psi'(\vec{p})\sigma(\vec{k}-\vec{p})-\psi(\vec{p})\psi(\vec{k}-\vec{p})-{1\over\mH^2}\psi'(\vec{p})\psi'(\vec{k}-\vec{p})+\(\vec{p}\cdot\vec{k}\)\sigma(\vec{p})\sigma(\vec{k}-\vec{p})\Big\}\no\\
&&=-4\int\frac{\mathrm{d}^3p}{(2\pi)^{3/2}} \Big(e^{ij}p_ip_j\Big)\Phi_p\Phi_{|\vec{p}-\vec{k}|}F_{S}(|\vec{p}|,|\vec{k}-\vec{p}|,\eta),
\n
Here, we put all the time dependent part into the transfer function which reads
\m
F_{S}(|\vec{p}|,|\vec{k}-\vec{p}|,\eta)=F_{S}(u,v,x)={3u\over v}T_\psi'(ux)T_\sigma(vx)-T_\psi(ux)T_\psi(vx)-uvx^2T_\psi'(ux)T_\psi'(vx)+{1+u^2-v^2\over 2uv}T_\sigma(ux)T_\sigma(vx).\no\\
\n
The dimensionless variables are defined as $u\equiv p/k$, $v\equiv |\vec{p}-\vec{k}|/k$ and $x\equiv k\eta$.
Our convention for the polarization tensor results in $e^{ij}p_ip_j=p^2/\sqrt{2}\sin^2\theta\cos2\phi$ and $p^2/\sqrt{2}\sin^2\theta\sin2\phi$ for $+$ and $\times$ modes respectively. 
Unless otherwise being stated, the prime with $T'(y)$ represents the derivative with respect to $y$, other than the conformal time $\eta$.
By defining the two-point correlator as
\e
\<C_kC_{k'}\>\equiv {2\pi^2\over k^3}P_C(k)\delta(\vec{k}+\vec{k'}),
\q
the density parameter which is defined by the energy density per logarithm wavelength normalized by the critical energy density can be evaluated as (see \eg \cite{Kohri:2018awv})
\m
\ogw^{S}(k)&&= \frac{1}{24} \(\frac{k}{\mH}\)^2 \overline{P_h(k)}= \(\frac{k}{\mH}\)^2{k^3\over48\pi^2{a(\eta)}^2}\int\mathrm{d}\tilde{\eta_1}\mathrm{d}\tilde{\eta_2}g_k(\eta;\tilde{\eta_1})g_{k'}(\eta;\tilde{\eta_2})
a({\tilde{\eta_1}})a(\tilde{\eta_2})\<{S}(\vec{k},\tilde{\eta_1}){S}(\vec{k'},\tilde{\eta_2})\>.\no\\
&&=\(\frac{k}{\mH}\)^2{4\pi^2k^3\over3a(\eta)^2}\int\frac{\mathrm{d}^3p}{(2\pi)^3}
\int\mathrm{d}\tilde{\eta_1}\mathrm{d}\tilde{\eta_2}{a(\tilde{\eta_1})a(\tilde{\eta_2})}g_k(\eta;\tilde{\eta_1})g_k(\eta;\tilde{\eta_2})
\Big(e^{ij}p_ip_j\Big)^2{1\over p^3|\vec{k}-\vec{p}|^3}P_\Phi(k)P_\Phi(|\vec{k}-\vec{p}|)\no\\
&&\qquad\qquad\qquad\qquad\qquad\qquad\times\Big[{F}_{S}(|\vec{p}|,|\vec{k}-\vec{p}|,\tilde{\eta_1}){F}_{S}(|\vec{p}|,|\vec{k}-\vec{p}|,\tilde{\eta_2})+{F}_{S}(|\vec{p}|,|\vec{k}-\vec{p}|,\tilde{\eta_1}){F}_{S}(|\vec{k}-\vec{p}|,|\vec{p}|,\tilde{\eta_2})\Big]\no\\
&&=\(\frac{k}{\mH}\)^2{8\pi^2k\over3}\int\frac{\mathrm{d}^3p}{(2\pi)^3}\(\int\mathrm{d}\tilde{\eta_1}{a(\tilde{\eta_1})\over a(\eta)}kg_k(\eta;\tilde{\eta_1})\tilde{F}_{S}(|\vec{p}|,|\vec{k}-\vec{p}|,\tilde{\eta_1})\)^2
\Big(e^{ij}p_ip_j\Big)^2{1\over p^3|\vec{k}-\vec{p}|^3}P_\Phi(k)P_\Phi(|\vec{k}-\vec{p}|)\no\\
\label{gw}&&={1\over 6}\int_0^\infty\mathrm{d}u\int_{|1-u|}^{1+u}\mathrm{d}v~{v^2\over u^2}\Big[1-\({1+v^2-u^2\over 2v}\)^2\Big]^2P_\Phi(uk)P_\Phi(vk)\overline{I_{S}^2(u,v,x\rightarrow\infty)},
\n
where we have used that $\mH=\eta^{-1}$ and $a(\tilde{\eta})/a(\eta)=\tilde{\eta}/\eta$ in RD era and we have summed over the two polarization modes. Eq.~(\ref{gw}) is the density parameter evaluated at matter-radiation equality. To get the density parameter by today, one needs to multiply the radiation density parameter, $\Omega_{\rm{r}}$. In the last step of Eq.~(\ref{gw}), we define
\m\label{is}
I_{S}(u,v,x)&&\equiv\int_0^x\mathrm{d}\tx~\tx\sin(x-\tx)\tilde{F}_S(u,v,x),
\n
with the overline denotes the oscillating average and $\tilde{F}_S(u,v,x)\equiv (F_S(u,v,x)+F_S(v,u,x))/2$ is the symmetric part of the transfer function. Although one may write the transfer function in an unsymmetric way, it is equivalent to extract its symmetric part and perform the calculation. We follow \cite{Kohri:2018awv} to obtain the analytical expression for $\overline{I^2_{S}}$. The indefinite integral of Eq.~(\ref{is}) is given by
\m
I_S(u,v,x)&&=-{27(u^2+v^2-3)^2\over 16u^3v^3}
\Bigg\{
\Bigg[
\Si\(\(1-{(u+v)\over\sqrt{3}}\)x\)
+
\Si\(\(1+{(u+v)\over\sqrt{3}}\)x\)
-
\Si\(\(1+{(u-v)\over\sqrt{3}}\)x\)\no\\
&&
-\Si\(\(1-{(u-v)\over\sqrt{3}}\)x\)
\Bigg] \cos x
+\Bigg[
\Ci\(\(1+{(u-v)\over\sqrt{3}}\)x\)
+
\Ci\(\(1-{(u-v)\over\sqrt{3}}\)x\)
-
\Ci\(\Bigg|1-{(u+v)\over\sqrt{3}}\Bigg|x\)\no\\
&&
-
\Ci\(\(1+{(u+v)\over\sqrt{3}}\)x\)
\Bigg]\sin x
+{1\over u^2+v^2-3}\(4uv+(u^2+v^2-3)\ln\Bigg|
1-\frac{4uv}{(u+v)^2-3}
\Bigg|\)\sin x\no\\
&&+{12\over x^2(u^2+v^2-3)^2}\Bigg[
2u\cos{ux\over\sqrt{3}}\(vx\cos{vx\over\sqrt{3}}-\sqrt{3}\sin{vx\over\sqrt{3}}\)+2u\(-vx+\sqrt{3}\sin{vx\over\sqrt{3}}\)\no\\
&&+\sin{ux\over\sqrt{3}}\(
2\sqrt{3}v-2\sqrt{3}v\cos{vx\over\sqrt{3}}+(u^2+v^2-3)x\sin{vx\sqrt{3}}
\)
\Bigg]
\Bigg\}
\n

Then the expression of $\overline{I^2_S}$ takes the form
\m\label{I2S}
\overline{I_{S}^2}&&=\frac{729(u^2+v^2-3)^2}{512u^6v^6}\Bigg\{\Big(-4uv+(u^2+v^2-3)\ln\Big|{3-(u+v)^2\over 3-(u-v)^2}\Big|\Big)^2+\pi^2\(u^2+v^2-3\)^2\Theta(u+v-\sqrt{3})\Bigg\}.
\n
Here, we use the Bardeen potential $\Phi$ to calculate the SIGWs while the authors in \cite{Kohri:2018awv} used the comoving curvature $\zeta$ which is related to $\Phi$ by $\zeta=(3/2)\Phi$.

{\it SIGWs in Newton gauge. }
The conformal Newton gauge, which is also known variously as orthogonal zero-shear gauge, Poison gauge or longitudinal gauge, demands that $B=C=0$. Under Newton gauge, $\phi=\psi=\Phi$ is just the usual gauge-invariant gravitational potential. This gauge has been well studied over the past decade and we only take a brief review. The source term is given by
\m
S_{ij}^{N}=3\phi\p_i\p_j\phi-{2\over\mH}\p_i\phi'\p_j\phi-{1\over\mH^2}\p_i\phi'\p_j\phi'.
\n
The transfer function in Newton gauge is
\m
\tilde{F}_{N}(u,v,x)&&=3T_\phi(ux)T_\phi(vx)+uxT_\phi'(ux)T_\phi(vx)+vxT_\phi'(vx)T_\phi(vx)+uvx^2T_\phi'(ux)T_\phi'(vx),
\n
and $T_\phi$ can be easily obtained since $\phi=\Phi$, namely
\m\label{tphi}
\phi(\vec{k})=\Phi(\vec{k})&&\equiv\Phi_k
T_\Phi(k\eta)=\Phi_k{9\over x^2}\({\sin(x/\sqrt{3})\over x/\sqrt{3}}-\cos(x/\sqrt{3})\).
\n
A simple way to derive this result is to construct $\Phi$ through Eq.~(\ref{Phi}) in synchronous gauge with $\psi$ and $\sigma$ given by Eq.~(\ref{psisigma}).
Following the procedure in synchronous gauge, $\overline{I^2}$ can be evaluated as \cite{Espinosa:2018eve,Kohri:2018awv}
\m\label{I2N}
\overline{I_{N}^2(u,v,x\rightarrow\infty)}=\overline{I_{S}^2(u,v,x\rightarrow\infty)}.
\n

{\it SIGWs in uniform curvature gauge. }
Another choice of gauge is the so-called uniform curvature gauge by setting $\psi=C=0$. The SIGWs in this gauge was studied recently in \cite{Tomikawa:2019tvi}. The source term in uniform curvature gauge is given by
\m
S^{U}_{ij}=(\p^2B)(\p_i\p_jB)-(\p_b\p_iB)(\p^b\p_jB)+4\mH\phi\p_i\p_jB+\phi'\p_i\p_jB+2\phi\p_i\p_jB'+2\phi\p_i\p_j\phi,
\n
Although a standard way is to derive the equation of motion for $\phi$ and $B$, we can express the source term by $\Phi$ based on Eq.~(\ref{Phi}). In this way, one can write the transfer function as
\m
\tilde{F}_{U}(u,v,x)={1\over 6}(3-u^2-v^2)x^2T_\Phi(ux)T_\Phi(vx),
\n
without solving $\phi$ and $B$. $T_\Phi$ is given by Eq.~(\ref{tphi}), and $\overline{I^2}$ can be obtained as
\m\label{I2U}
\overline{I_{U}^2(u,v,x\rightarrow\infty)}=\overline{I_{S}^2(u,v,x\rightarrow\infty)}=\overline{I_{N}^2(u,v,x\rightarrow\infty)}.
\n
In \cite{Tomikawa:2019tvi}, the authors analytically calculated $\overline{I^2}$ for $w=1/3$ and $w=0$ and they numerically checked that $\overline{I^2}$ is identical in Newton gauge and uniform curvature gauge for any $w>0$.

{\it Result of $\ogw(k)$. }
In general, the energy density spectrum of SIGWs at matter-radiation equality is given by 
\e\label{ogw}
\ogw(k)
={1\over 6}\int_0^\infty\mathrm{d}u\int_{|1-u|}^{1+u}\mathrm{d}v~{v^2\over u^2}\Big[1-\({1+v^2-u^2\over 2v}\)^2\Big]^2P_\Phi(uk)P_\Phi(vk)
\overline{I^2(u,v,x\rightarrow\infty)}.
\q
Choosing a gauge will fix the expression of $\overline{I^2(u,v,x\rightarrow\infty)}$. Since $\overline{I^2}$ is identical in these three gauge, they will generate the same density parameter.

\begin{figure}[htbp!]
	\centering
	\includegraphics[width = 0.48\textwidth]{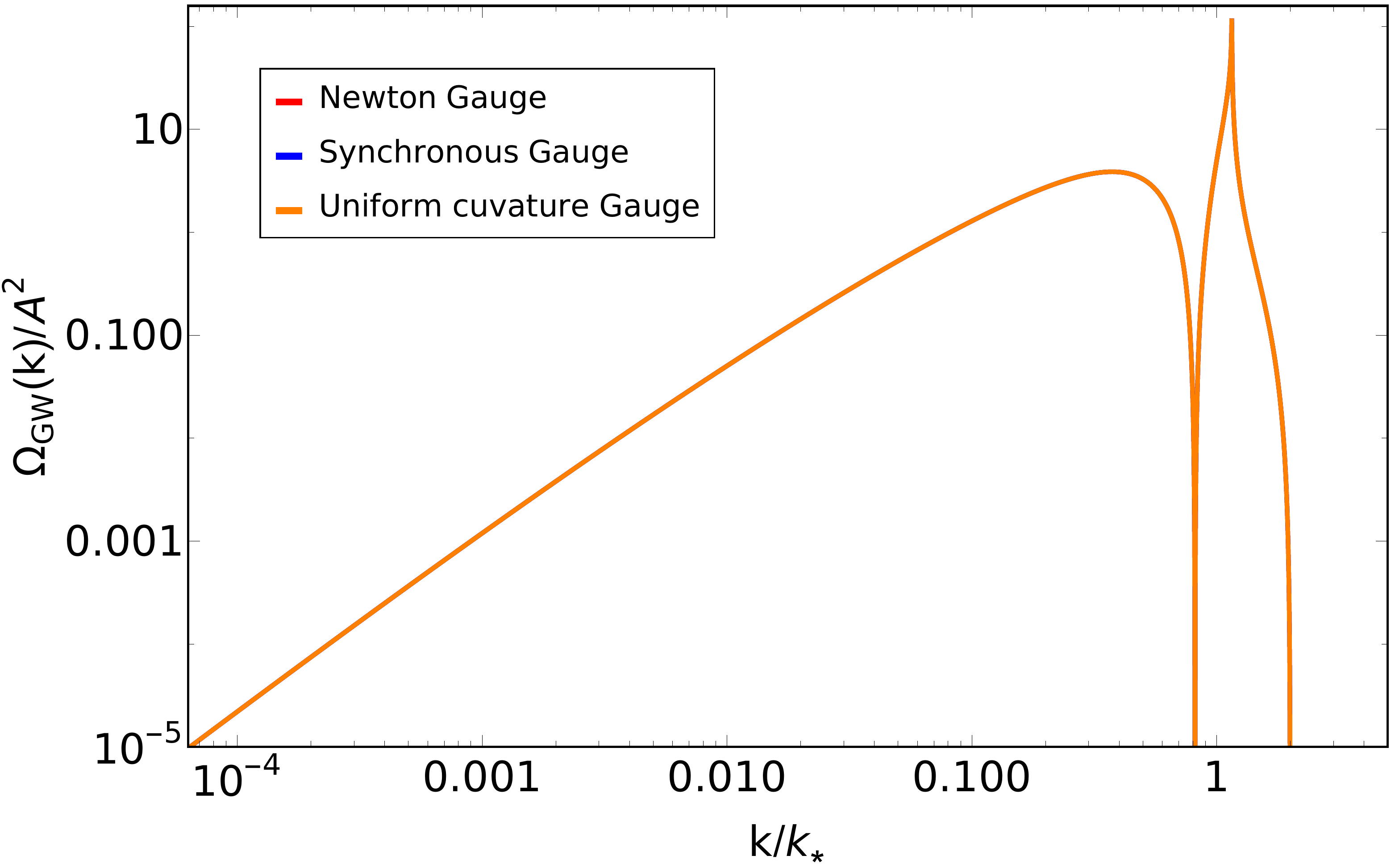}
	\includegraphics[width = 0.48\textwidth]{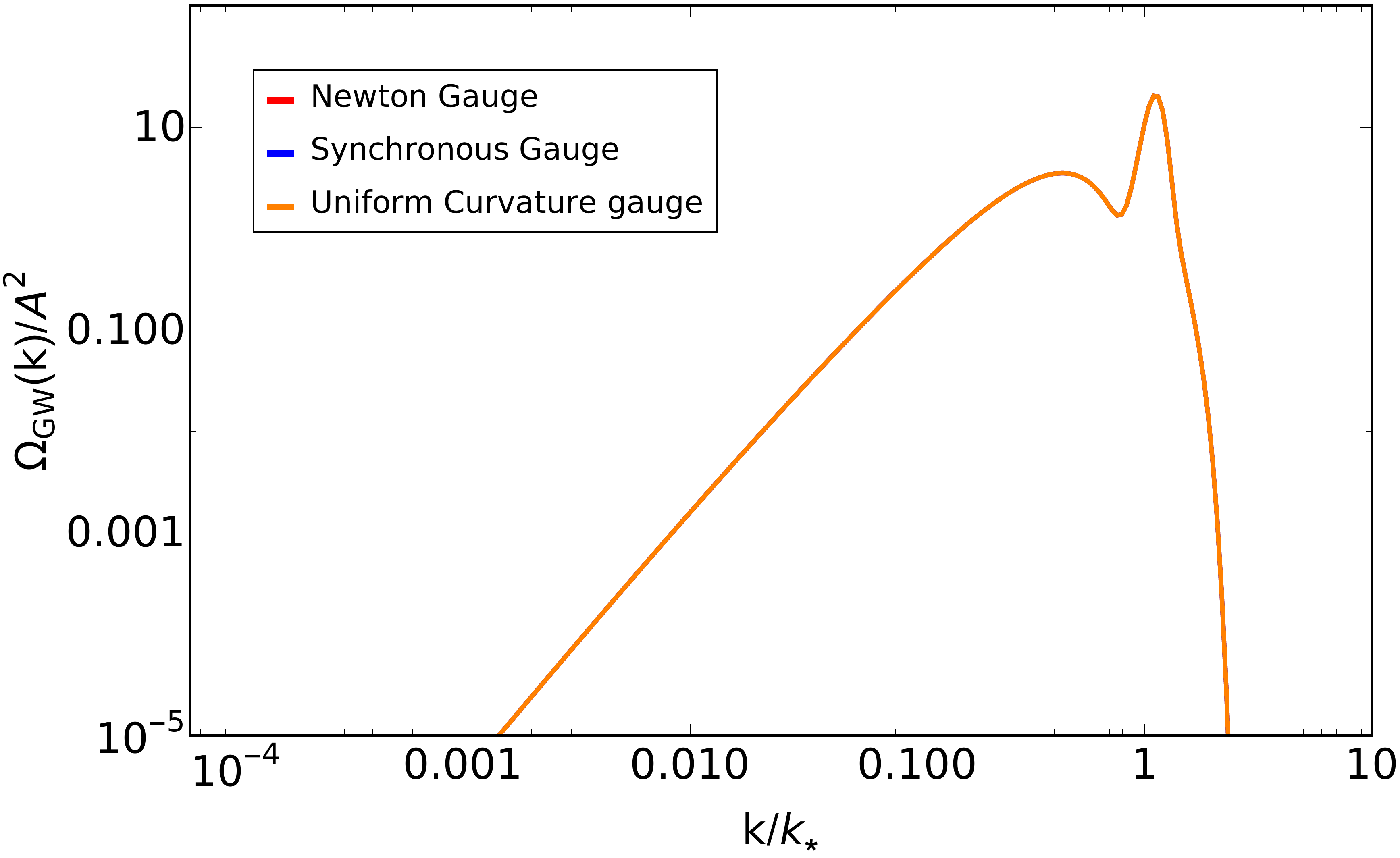}
	\caption{\label{gaussianogw} The density parameter of SIGWs in different gauges generated by a $\delta$-power spectrum (left panel) and a Gaussian-like power spectrum (right panel). The width of the Gaussian-like spectrum is set as $\sigma_*/k_*=0.1$.
	}
\end{figure}

First of all, we consider a $\delta$-power spectrum parameterized by $P_\Phi(k)=Ak_*\delta(k-k_*)$, with a dimensionless amplitude $A$ peaked at $k_*$. A $\delta$-power spectrum corresponds to a monochromatic PBH formation and is very useful in analytical studies. The density parameter is exactly the same and has the following functional form
\e
\ogw(k)={243 A^2\over 1024}\tilde{k}^2\(1-{\tilde{k}^2\over4}\)^2(2-3\tilde{k}^2)^2\Theta(2-\tilde{k})\Bigg((2-3\tilde{k}^2)^2\pi^2\Theta(2-\sqrt{3}\tilde{k})+\(-4+(2-3\tilde{k}^2)\ln\Big|1-{4\over 3\tilde{k}^2}\Big|\)^2\Bigg),
\q
where $\tilde{k}\equiv k/k_*$ is the dimensionless wavelength.
One can also consider a Gaussian-like power spectrum parameterized by
\m
P_\Phi(k)={Ak_*\over \sqrt{2\pi}\sigma_*}\exp\(-{(k-k_*)^2\over2\sigma_*^2}\).
\n
Our numerical results for the density parameter $\ogw(k)$ in different gauges are illustrated in Fig.~\ref{gaussianogw}. 

{\it Summary and discussion. }
In this letter we explicitly calculate the energy density spectra of SIGWs, $\Omega_{\rm{GW}}(k)$, in Newton gauge, synchronous gauge and uniform curvature gauge, respectively, and we find that the energy density spectra in different gauges have the same form. It implies that the energy density spectrum of SIGW may be gauge independent. 
Again, referring to \cite{Yuan:2019wwo}, the log-dependent slope of $\ogw(k)$ in the infrared region can be taken as a distinguishing feature for the SIGWs.

{\it Note added.}
After \cite{DeLuca:2019ufz} appeared in arXiv, Inomata and Terada revisited the SIGWs during radiation dominated era in both Newton gauge and synchronous gauge \cite{Inomata:2019yww} in which they found that the density parameter is identical in these two gauges. Recently, the authors in \cite{DeLuca:2019ufz} corrected their results in v1 and obtained the same conclusion. 

{\it Acknowledgments.}
We thank the anonymous referee for providing constructive comments and suggestions to improve the quality of this paper.
Cosmological perturbations are derived using the \texttt{xPand} \cite{Pitrou:2013hga} package. We also acknowledge the use of HPC Cluster of ITP-CAS. 
This work is supported by grants from NSFC (grant No. 11690021, 11975019, 11847612, 11991053), the Strategic Priority Research Program of Chinese Academy of Sciences (Grant No. XDB23000000, XDA15020701), and Key Research Program of Frontier Sciences, CAS, Grant NO. ZDBS-LY-7009.     

\bibliography{./TTgaugeref}
	
\end{document}